\documentclass[fontsize=11pt,a4paper]{scrartcl}

\addtokomafont{disposition}{\rmfamily}

\usepackage[T1]{fontenc}
\usepackage[utf8]{inputenc}

\usepackage{lineno}
\usepackage{csquotes}
\usepackage[english]{babel}
\usepackage[authordate,backend=biber]{biblatex-chicago}
\usepackage{hyperref}

\usepackage{graphicx}
\usepackage{authblk}
\usepackage[a4paper]{geometry}
\usepackage{lmodern}

\usepackage{color}

\title{How are research data referenced? The use case of the research data repository RADAR}

\author[1]{Dorothea Strecker (corresponding author; dorothea.strecker@hu-berlin.de)}
\author[2]{Kerstin Soltau}
\author[2]{Felix Bach}
\affil[1]{Berlin School of Library and Information Science, Humboldt-Universität zu Berlin}
\affil[2]{FIZ Karlsruhe - Leibniz Institute for Information Infrastructure}

\begin{document}
\maketitle

\section*{ORCID}
\begin{itemize}
    \item Dorothea Strecker: \url{https://orcid.org/0000-0002-9754-3807}
    \item Kerstin Soltau: \url{https://orcid.org/0000-0002-6368-1929}
    \item Felix Bach: \url{https://orcid.org/0000-0002-5035-7978}
\end{itemize}

\section*{Keywords}
\textbf{Keywords:} research data ; data citation ; research data repository ; data reuse ; data referencing

\section*{Abstract}
Publishing research data aims to improve the transparency of research results and facilitate the reuse of datasets. In both cases, referencing the datasets that were used is recommended. Research data repositories can support data referencing through various measures and also benefit from it, for example using this information to demonstrate their impact. However, the literature shows that the practice of formally citing research data is not widespread, data metrics are not yet established, and effective incentive structures are lacking.
This article examines how often and in what form datasets published via the research data repository RADAR are referenced. For this purpose, the data sources Google Scholar, DataCite Event Data and the Data Citation Corpus were analyzed.
\\
The analysis shows that 27.9 \% of the datasets in the repository were referenced at least once. 21.4 \% of these references were (also) present in the reference lists and are therefore considered \emph{data citations}. Datasets were referenced often in data availability statements. A comparison of the three data sources showed that there was little overlap in the coverage of references. In most cases (75.8 \%), data and referencing objects were published in the same year. Two definition approaches were considered to investigate \emph{data reuse}. 118 RADAR datasets were referenced more than once. Only 21 references had no overlaps in the authorship information -- these datasets were referenced by researchers that were not involved in data collection.

\section{Introduction}
The practice of citing datasets in research publications has been promoted for many years now. The Force 11 Joint Declaration of data citation principles state that “[...] data should be considered legitimate, citable products of research. Data citation, like the citation of other evidence and sources, is good research practice and is part of the scholarly ecosystem supporting data reuse.” \parencite{data_citation_synthesis_group_joint_2014} Data citation is ascribed many positive effects, including the attribution of data creators, the positioning of datasets within the scholarly record, or the assessment of the impact of datasets \parencite{silvello_theory_2018}.
\\
Research data repositories are important actors in realizing the vision of habitual data citation. They are facilitators of data citation and in turn might be interested in data metrics to communicate their impact \parencite{lowenberg_open_2019,puebla_grei_2024}.
Data sharing, access and reuse are common themes in literature on “long-tail” research data - “[v]ast amounts of distributed, heterogeneous, often smaller scientific datasets of various types generated by individual researchers or small research groups; these data do not receive the same level of attention, funding, or infrastructure support as larger, well managed datasets but nevertheless contain a wealth of valuable specialized information that can potentially contribute to scientific knowledge.” \parencite[p. 3]{stahlman_evolution_2024}
\\
This paper discusses the concepts \emph{data use} and \emph{data referencing}, outlines the measures a research data repository specialized on long-tail research data, RADAR\footnote{RADAR: \url{https://doi.org/10.17616/R3ZX96}; \textit{Last accessed on May 13th, 2025.}}, has implemented to support data referencing, and investigates how the collection of RADAR is referenced.

\subsection{Data use}
Researchers are advised to cite data when reusing them \parencite{data_citation_synthesis_group_joint_2014}. This recommendation makes the process and motives of data referencing sound simple, and suggests a direct connection between using and referencing data. However, previous research has shown that research practices are complex, and the concepts referencing and using data need to be differentiated further to gain a better understanding of what \emph{data citation} means.
\\
Previous research has shown that data are used for varying purposes; some types of data use don't result in new findings (\emph{background use}), and researchers therefore might not consider indicating data use in publications \parencite{wynholds_data_2012,banaeefar_best_2022}.
Data reuse is often distinguished from data use as an essential variant. However, upon closer inspection, the concept \emph{data reuse} is not clearly defined. The literature discusses several approaches that could be used to identify data reuse:
\begin{itemize}
    \item A dataset is used by someone other than the original data creator (data reuse defined by authorship) \parencite{he_reuse_2016,pasquetto_reuse_2017}
    \item A dataset is used multiple times (data reuse defined by number of uses) \parencite{peters_research_2016}
    \item A dataset is used to answer new research questions (data reuse defined by purpose) \parencite{sandt_definition_2019}
\end{itemize}
\cite{sandt_definition_2019} argue that all of these approaches have shortcomings, and research practices do not always fall clearly into one of the two categories \emph{data use} and \emph{data reuse}.

\subsection{Data referencing}
Researchers choose different methods to refer to data. \cite{gregory_tracing_2023} categorized referencing practices into \emph{data citation} (data are referenced in the reference list), \emph{data mention} (data are referenced in other parts of the publication), and \emph{indirect data citation} (a related publication is referenced in the reference list). Previous research showed that data citation is rare, and that authors often choose other methods of referencing data \parencite{belter_measuring_2014,park_examination_2017,park_informal_2018,quarati_researchers_2022}. Bibliographic databases also indicate that data citation remains rare, with only a small proportion of datasets being cited overall \parencite{ninkov_datasets_2021,peters_research_2016,robinson-garcia_analyzing_2016}. However, there is some evidence that data citation increased over time \parencite{peters_research_2016,he_reuse_2016}. In comparison to evidence from bibliographic databases, authors also report higher rates of referencing data: In a survey,  58.3 \% of participants reported that they often or always cite or mention data they use \parencite{gregory_tracing_2023}. Studies that consider data citations as well as data mentions reveal higher rates of data referencing \parencite{belter_measuring_2014,gerasimov_comparison_2024}. There is also evidence that indirect data citation is common. 38 \% of data papers - documents describing datasets and their generation - were cited \parencite{stuart_data_2017}. The analysis of a dataset from cancer research showed that indirect citation was more common than citing the dataset directly \parencite{yoon_how_2019}. This suggests that bibliographic databases don't capture all forms of referencing data.
\\
Characteristics of datasets impact the number of citations they receive. The number of citations Dryad datasets accrue in Scopus varies by discipline, and less aggregated forms were generally more likely to be cited \parencite{he_reuse_2016}. An analysis of the Data Citation Index (DCI) resulted in similar findings \parencite{peters_research_2016}. This highlights the need to contextualize citation analyses by considering the subject of a dataset; however, metadata describing research data don't always include that information \parencite{he_reuse_2016,ninkov_datasets_2021}.
\\
There are many challenges to detect and collect references to data, starting with methods authors choose to refer to datasets. Research has shown that when researchers refer to datasets, they sometimes refer to research data repositories instead \parencite{peters_research_2016,yang_research_2025}, or refer to datasets by name \parencite{mathiak_challenges_2015}. This makes it difficult to detect references and allocate them to specific datasets. Authors do not always use persistent forms of referencing datasets, such as DOIs \parencite{yoon_how_2019}. There are many more opportunities for data references to get lost, even if authors include them in documents, for example if they are not submitted to indexing services by publishers \parencite{borda_if_2023,lowenberg_open_2019}.
\\
Bibliographic resources vary significantly in the coverage they provide for data citations \parencite{gerasimov_comparison_2024}. Open citation data initiatives are perceived to have great potential in tracing data (re)use, but they face challenges such as reference mining in full texts or the reliance on publishers providing open references \parencite{taskin_sustaining_nodate}. The Data Citation Corpus aggregates references from several sources, including those that are not using DOIs \parencite{puebla_building_2024}.
\\
Several studies have analyzed self-citations, which can provide insights into the prevalence of data reuse. Bibliographic resources provide varying results when analyzed for rates of self-citation - the phenomenon is quite common in the DCI, but relatively low for datasets in OpenAlex \parencite{krause_who_2023,park_examination_2017}. This might be a result of different approaches to indexing of the two sources. Self-citation rates are very high for generalist repositories: \mbox{85 \%} of citations of Dryad datasets in Scopus and 98.5 \% of citations of Zenodo datasets were self-citations \parencite{he_reuse_2016,sandt_definition_2019}. These findings could indicate that generalist repositories are frequently used by authors to deposit their data as a means to satisfy data policies or guidelines by journals and other stakeholders.
\\
Repositories can facilitate data use. A study at the Inter-university Consortium for Political and Social Research (ICPSR), a social sciences data archive from the USA, has shown that an increase in the number of curatorial actions and subject terms assigned to data resulted in more users downloading data \parencite{hemphill_how_2022,hemphill_dataset_2024}. They might also be interested in using evidence of data use themselves to demonstrate their value to stakeholders, since data citation “signals the added value of data repositories” \parencite[p. 1]{puebla_grei_2024}.

\subsection{RADAR -- measures to support data use and referencing}
The cross-disciplinary RADAR repository from FIZ Karlsruhe - Leibniz Institute for Information Infrastructure was developed as part of a DFG project. It has been available to academic institutions since 2017 as a comprehensive cloud service for archiving, publishing and optionally reviewing research data. Since 2021, “RADAR Local” has also been available as an operating variant in which the software is run by FIZ Karlsruhe on institutional infrastructure. Furthermore, as a participant in the National Research Data Infrastructure (NFDI), FIZ Karlsruhe operates RADAR4Culture, RADAR4Chem, and RADAR4Memory - community-tailored publication services designed to meet the specific needs of their respective research communities. All RADAR offers and variants use the same software code. RADAR has already introduced a number of measures to promote the use and referencing of research data and to ensure that data remains usable in the long term.
\\
Each published dataset receives a DOI based on the RADAR metadata schema\footnote{RADAR Metadata Schema: \url{https://doi.org/10.25504/FAIRsharing.e26f92}; \textit{Last accessed on May 13th, 2025.}} with ten mandatory and 13 optional fields based on the DataCite Kernel 4.4. Links to ORCID, ROR or GND promote attribution or linking to standard data. At present, the integration of terminologies via the TS4NFDI\footnote{TS4NFDI: \url{https://ts4nfdi.github.io/homepage/}; \textit{Last accessed on May 13th, 2025.}} is under development. Digital resources or software can be related via the metadata which can also be updated after publication. The license for the data set is selected from a variety of different license options; metadata are published under CC0. On the landing page of each dataset, RADAR offers citation recommendations in seven citation styles as well as export formats such as BibTeX, RIS and EndNote.
\\
Particular attention is paid to the adherence to the FAIR Principles \parencite{wilkinson_fair_2016}. RADAR actively supports this through a variety of technical measures and even participated in the EU project FAIR-IMPACT. During this assessment process, the scores generated by the evaluator tool F-UJI \parencite{devaraju_fairsfair_2022}, an indicator for the adherence of data publications to the FAIR Principles, was significantly increased (up to 91 \%). RADAR metadata are widely available and machine-readable. Open interfaces such as REST API, OAI-PMH\footnote{RADAR OAI-PMH: \url{https://www.radar-service.eu/oai/}; \textit{Last accessed on May 13th, 2025.}} for harvesting and FAIR Signposting\footnote{FAIR Signposting: \url{https://signposting.org/FAIR/}; \textit{Last accessed on May 13th, 2025.}} are offered. Integration into a knowledge graph and access via a SPARQL endpoint also enable semantic integration of the RADAR datasets into higher-level research data ecosystems.
\\
Internal curation and an optional peer review step are provided for quality assurance during the publication workflow. Long-term archiving is implemented geo-redundantly in three copies at two locations. Published data is available for at least 25 years to guarantee potential for reuse over time.
\\
Statistics on the landing pages indicate the number of page views and data downloads. From the perspective of the repository operators, it would be desirable if this section could be supplemented with information on data citation and data reuse. An integrated comment function, the sharing of the dataset via social media or the embedding of landing pages on other websites could additionally promote the dissemination of RADAR datasets in the future.

\section{Method}
This study addresses the following questions:
\begin{enumerate}
    \item Do researchers reference RADAR datasets?
    \item Which methods do they choose when referencing RADAR datasets?
    \item How do data sources compare in terms of the coverage of references?
    \item How common is the reuse of RADAR datasets?
\end{enumerate}

\subsection{Data collection}
\begin{figure}[ht]
    \centering
    \includegraphics[width=0.9\textwidth]{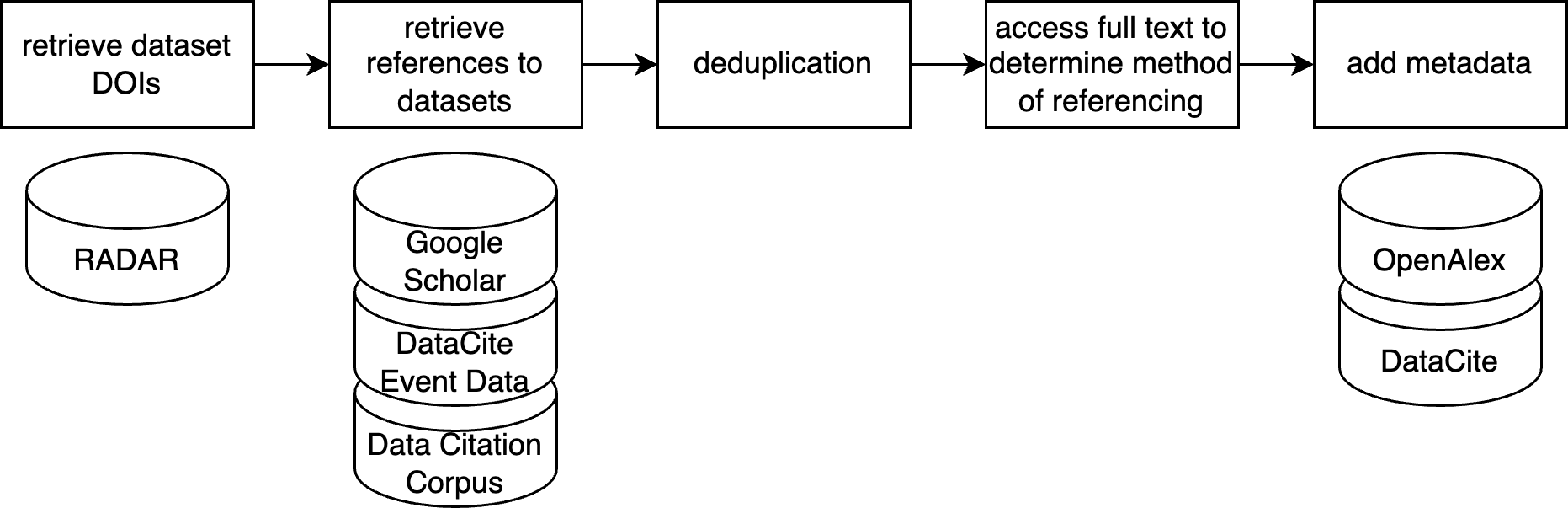}
    \caption{\small Outline of the data collection process.}
    \label{fig:figure_1}
\end{figure}

DOIs assigned to RADAR datasets were retrieved from the RADAR API 2025-01-27. In a second step, references attributable to these DOIs were collected from three data sources: Google Scholar, the Data Citation Corpus, and DataCite Event Data.
\\
Google Scholar indexes metadata and full texts of research outputs and was shown to offer comprehensive coverage of data references in previous research \parencite{gerasimov_comparison_2024}.
The Data Citation Corpus was developed by DataCite and the Chan Zuckerberg Initiative \parencite{vierkant_wellcome_2023}. It identifies data citations in various sources, including references in full texts \parencite{puebla_building_2024}.
DataCite Event Data captures relationships between research data and other objects that are manifested in metadata, using the property \emph{relatedIdentifier} of the DataCite Metadata Schema.\footnote{DataCite Event Data: \url{https://support.datacite.org/docs/consuming-citations-and-references}; \textit{Last accessed on May 13th, 2025.}}
These data sources were selected to cover different approaches of capturing references to research data. \cite{gerasimov_comparison_2024} compiled an overview of the scope and availability of additional data sources covering references to research data.
\\
References in Google Scholar were identified by searching for the dataset DOI in full texts between 2025-02-10 and 2025-02-12. All results were checked manually. Version 3.0 of the Data Citation Corpus was used, which was released 2025-02-01 \parencite{datacite_data_2025}. Event data was retrieved from the DataCite API 2025-01-27. The references obtained from the three sources were deduplicated.
Similar to the approach used by \parencite{khan_measuring_2021}, each full text referencing a RADAR dataset was accessed to determine where the reference occurred in the text (in the reference list, data availability statement, footnotes, full text; multiple methods of referencing were possible).
Metadata (author names, publication date) for datasets and referencing objects were added from OpenAlex and DataCite on 2025-02-10.
Figure \ref{fig:figure_1} gives an overview of the data collection process.
The resulting dataset includes 604 references between a RADAR dataset and a referencing object (DOI - DOI pairs). The data are published and can be accessed at RADAR \parencite{strecker_referenzierung_2025}.

\section{Results}

\subsection{References to RADAR datasets}
\begin{figure}[ht]
    \centering
    \includegraphics[width=0.9\textwidth]{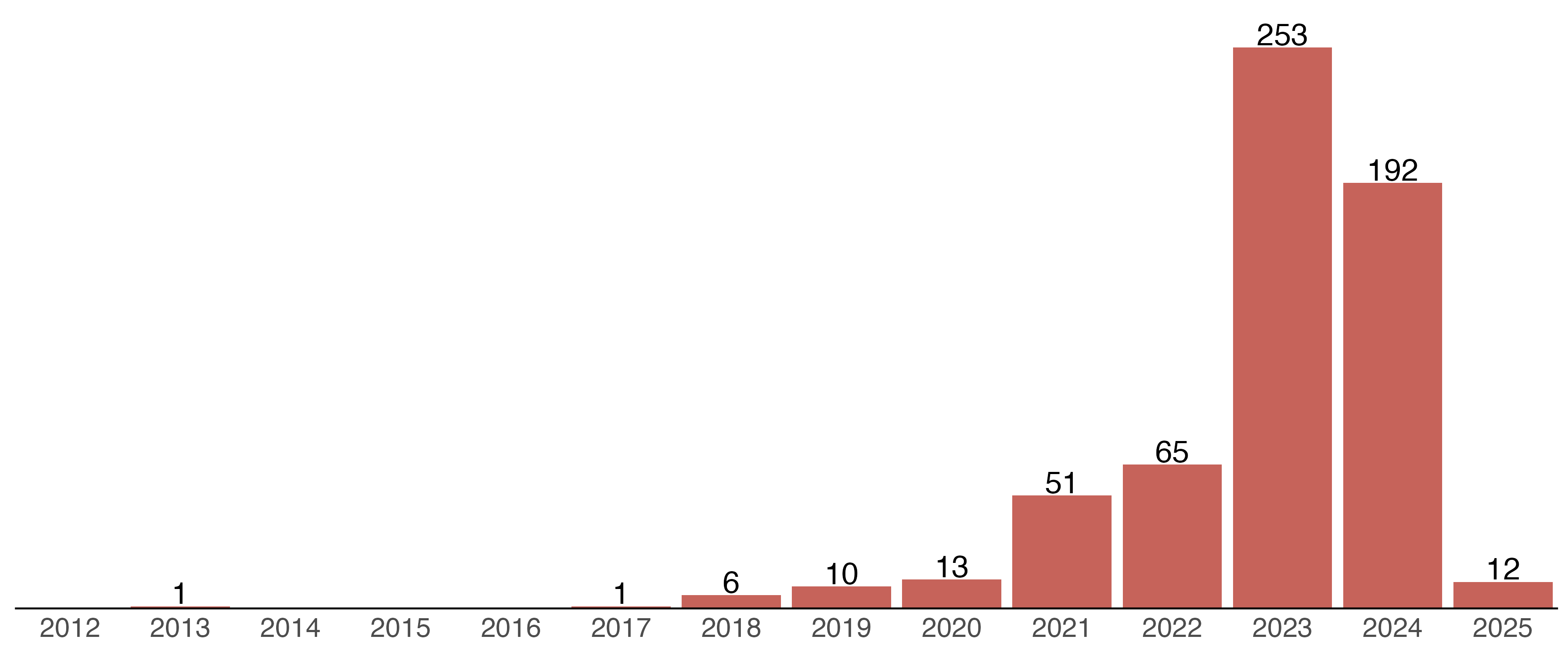}
    \caption{\small Distribution of the publication years of referencing objects.}
    \label{fig:figure_2}
\end{figure}

Of the 1.605 RADAR datasets, 27.9 \% (448) were referenced at least once. The objects referencing the datasets were published between 2013 and 2025. The number of publications increases from 2017, reaches a peak in 2023, and tapers off in the following years (see Figure \ref{fig:figure_2}).
In most cases (458 ; 75.8 \%), datasets and referencing objects were published in the same year. For the remaining cases, it was more common for data to be published before the referencing object (87 ; 14.4 \%) than in the reverse order (49 ; 8.1 \%).

\subsection{Referencing methods}
\begin{figure}[ht]
    \centering
    \includegraphics[width=0.9\textwidth]{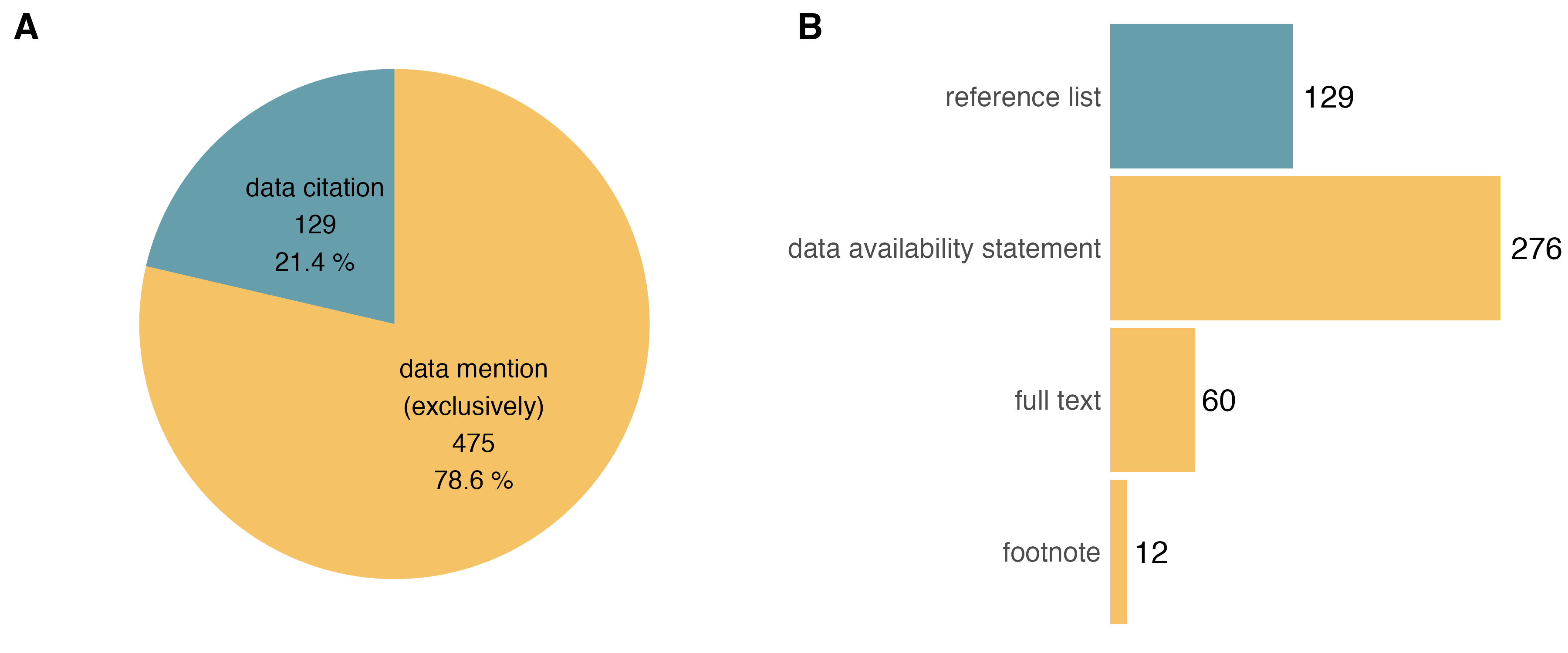}
    \caption{\small A referencing methods (based on \cite{gregory_tracing_2023}), n = 604 references ; B position of the reference in the referencing object (multiple options can apply).}
    \label{fig:figure_3}
\end{figure}

The analysis of the full text of the referencing object revealed that data citations (data are referenced in the reference list) represent 21.4 \% of the references in the sample, 78.6 \% are data mentions (data are referenced in other parts of the publication) (see Figure \ref{fig:figure_3}, A). This categorization is based on \parencite{gregory_tracing_2023}, and is described in more detail above.
\\
The most common position of data mentions (276) are in data availability statements, followed by other parts of the text, such as methods sections (60), and in footnotes (12) (see Figure \ref{fig:figure_3}, B). Authors use more than one of these referencing methods in 388 cases.
In 220 cases, no referencing method could be determined, either because the referencing object was not a text (for example datasets), or because the DOI was not found in the full text (for example because the citation style did not display the DOI).

\subsection{Coverage of data sources}
\begin{figure}[ht]
    \centering
    \includegraphics[width=0.9\textwidth]{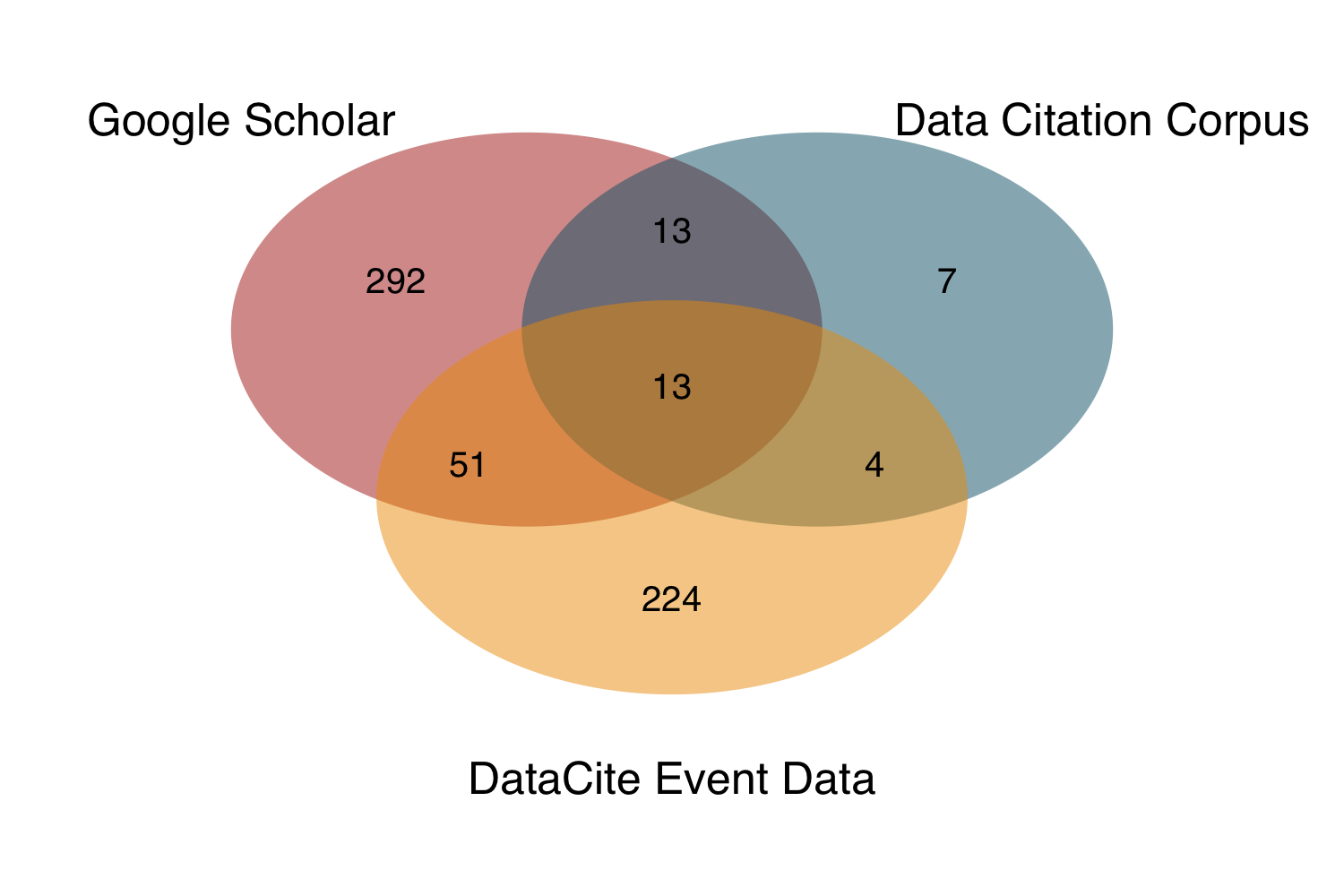}
    \caption{\small Number of references in the data sources.}
    \label{fig:figure_4}
\end{figure}

369 references to RADAR datasets were found in Google Scholar, 37 in the Data Citation Corpus and 292 in DataCite Event Data.
Some references were found in more than one data source, but overall, there is little overlap (see Figure \ref{fig:figure_4}). This applies to Google Scholar and DataCite Event Data in particular: Most references in the sample were identified in only one of these sources. Google Scholar covers most cases where datasets are referenced in the full text, whereas DataCite Event Data covers most cases where references are established in metadata only.

\subsection{Reuse of RADAR datasets}
Two definitions were used to determine whether a RADAR dataset was reused: 1) the dataset was referenced more than once, and 2) there is no overlap in authorship between the dataset and the referencing object. 118 datasets in the sample (7.4 \% of all RADAR datasets) were referenced more than once and could be considered “reused”. However, it is important to note that there are preprints and corrections among these referencing objects that skew the results. 21 datasets (1.3 \% of all RADAR datasets) have no overlap in authorship, which means they are referenced by researchers that were not involved in data collection.

\section{Discussion}
The analysis shows that RADAR datasets are referenced - a considerable share of the datasets published in RADAR were referenced at least once. The results also indicate that the practice of publishing and referencing datasets might have become more common over time.
\\
“Data  citation” can be a useful metaphor, for example to communicate to researchers the relevance of publishing and referencing their data for ensuring transparency of results. However, referencing data differs from referencing literature in some key aspects. For example, the analysis has shown that researchers often use different methods for referencing data, and self-citation is very common. This could indicate that researchers attribute different meanings to referencing data. More research is needed to investigate what these references mean in the scholarly ecosysten; in the meantime, the metaphor “data citation” should be used carefully.
Future research could also investigate whether references to research data vary depending on the type of repository they are published in. The self-citation rate at RADAR is very high (98.7 \%). This observation is consistent with studies of other generalist repositories that are outlined above. It remains to be determined if this is the case for all types of research data repositories, or generalist repositories specifically.
\\
The analysis also revealed that data sources vary significantly in terms of the references they cover. This is likely a result of researchers’ referencing practices and the workflows data sources have established for recording connections between data and text publications. It is particularly noteworthy that in the case of RADAR, data sources tend to either capture references made in the full text or references made in metadata. To get a more complete representation of references to data, analyses should use multiple sources.
\\
Although researchers’ motives for referencing data were not the primary focus of this study, the results indicate that researchers reference data to adhere to policies, for example of journals or funders. This assumption seems plausible, because the most prevalent method researchers chose when referencing data was data availability statements, which are sometimes mandated or recommended to them. Unfortunately, adhering to these policies does currently not lead to the promised incentives for data sharing in the form of “impact” actually being realized, because the recommended referencing methods (data availability statements) are not recorded by databases that are easily machine readable. To make data referencing rewarding for researchers, policies and workflows of sources for reference data should be aligned more closely.
\\
Repositories that want to use reference data to communicate the impact of datasets in their collection to stakeholders need a complete, accurate, and convenient source for this type of data. At the moment, reference data is dispersed and has to be reviewed and cleaned manually before it can be used. A truly useful data source would also require the cooperation of multiple stakeholders to ensure that references are counted as completely as possible.
However, reference data alone do not convey impact. Repositories need data metrics - careful interpretations of reference data that give meaningful insights. This includes clear definitions of concepts like data use and data citation, and a consistent system for counting references. This is one of the goals of a new FORCE 11 group, “Data usage typologies”\footnote{FORCE 11 group “Data usage typologies": \url{https://force11.org/group/data-usage-typologies/}; \textit{Last accessed on May 13th, 2025.}}. Overall, data metrics must be developed carefully to avoid harmful effects and misuse, as seen with metrics like the Journal Impact Factor.

\section{Conclusion}
This study demonstrated that datasets published in RADAR are referenced in journal articles and other research outputs. High self-citation rates and references in data availability statements suggest that researchers primarily reference data to make results of their own research more transparent, not because they reused existing data. Although authors might be asked to reference data in data availability statements, these references are often missing from sources of reference data. Policies and workflows of sources for reference data should be aligned more closely to reward data referencing.
\\
The process also showed that analyses of data citations currently require a considerable amount of effort and care. Sources for reference data vary significantly in terms of their coverage and often require manual cleaning. Reference counts should be interpreted with caution - in order to meaningfully communicate their impact, repositories need data metrics, which are still under development.

\section*{Authorship Contributions}
\begin{itemize}
    \item [] \textbf{Dorothea Strecker} conceptualization ; writing (original draft) ; formal analysis ; data curation
    \item [] \textbf{Kerstin Soltau} conceptualization ; writing (original draft)
    \item [] \textbf{Felix Bach} conceptualization ; writing (original draft)
\end{itemize}

\section*{Conflict of Interest}
The authors have no conflicts of interest to declare.

\printbibliography

\end{document}